\documentclass[letterpaper,aps,prl,twocolumn,showpacs,superscriptaddress]{revtex4}

\usepackage{graphicx}
\usepackage{bm}
\usepackage{amsmath}

\begin{document}
\title{Generation of spin currents via Raman scattering}
\author{Ali Najmaie}
\email{anajmaie@physics.utoronto.ca}
\author{E. Ya. Sherman}
\author{J. E. Sipe}
\address{Department of Physics and Institute for Optical Sciences, University of Toronto, 60 St.
George St., Toronto, Ontario, Canada M5S 1A7}
\date{\today}

\begin{abstract}
We show theoretically that stimulated spin flip Raman scattering
can be used to inject spin currents in doped semiconductors with
spin split bands. A pure spin current, where oppositely oriented
spins move in opposite directions, can be injected in zincblende
crystals and structures. The calculated spin current should be
detectable by pump-probe optical spectroscopy and anomalous Hall
effect measurement.
\end{abstract}


\maketitle

Spintronics is the strategy of using the spins of carriers in
solid state structures as a new degree of freedom to carry and
transform information \cite{opticalO}. An important capability is
the generation of spin currents, which couple spin and charge.
Suggestions for the generation of spin currents have included
proposals based on the use of the extrinsic and intrinsic spin
Hall effects, as well as spin pumping
\cite{KatoetalSHE,WunderlichetalSHE,Watsonetal}. All-optical
generation and control of spin currents have also been proposed
and observed \cite
{Hubneretal,Stevensetal,Bhatetal,Shermanetal,Ganichevetal}. In
these optical schemes, the generation of the spin current is
accompanied by the deposition of photon energies of about 1 eV per
carrier in the system. Most of this energy is used simply to
generate the electron-hole pairs, and is lost in the sense that it
does not contribute to the motion of the carriers. An obvious goal
for the all-optical manipulation of spin is the minimization of
such energy loss.

Here we propose a new scheme for generating spin currents in doped
semiconductors via a two photon Raman scattering process \cite
{Jusserandetal,LightScattering}, which can be understood as the
absorption of a photon of frequency $\omega _{1}$ and the emission
of a photon at $ \omega _{2}$. This process deposits an energy of
$\hbar \Omega =\hbar \left( \omega _{1}-\omega _{2}\right) $ per
carrier in the crystal, which for GaAs is only about 0.5 meV. In
the presence of spin splitting of the bands, stimulated spin-flip
Raman scattering can produce a pure spin current, where electrons
with opposite spins move in opposite directions. Hence no net
charge current is injected.

Crucial to this effect is the spin-splitting of states, which
results from the lack of inversion symmetry of the crystal. A
schematic picture of the conduction states is shown in Fig. 1,
where $E_{F}$ denotes the Fermi energy. For simplicity we take the
temperature to be zero \cite{footnote}. The proposal in this
letter is based on optically inducing transitions between states
$d(\pm {\bf k)}$ and $u(\pm {\bf k})$ at $\pm {\bf k}$
wavevectors, as depicted in Fig. 1, via Raman scattering.
\begin{figure}
\includegraphics[width=5.0cm]{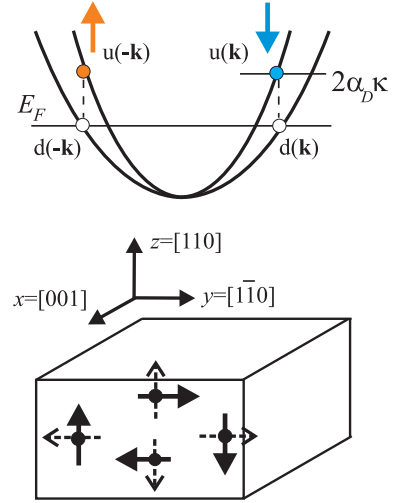}
\caption{(Color online) Schematic picture of the spin-split
conduction bands. Spin-flip Raman scatterings correspond to
transitions from $d(\pm {\bf k)}$ to $u(\pm {\bf k})$ at $\pm {\bf
k}$. The carriers at $u(\pm {\bf k})$ then have opposite spin
orientations and hence a pure spin current is injected in the
system. An example of the generated spin current is presented
where two pure spin currents with orthogonal velocities are shown
schematically. The dotted lines correspond to the velocities of
the carriers, and the thick arrows depict their spins.}
\end{figure}
As an example, we first consider an optical field ${\bf E}={\bf
E}(\omega _{1})e^{-i\omega _{1}t}+{\bf E}(\omega _{2})e^{-i\omega
_{2}t}+c.c.$ in the material, where $c.c.$ denotes complex
conjugation and $\omega _{1}>\omega _{2}$ and then present an
experimentally relevant scenario of pulse excitation. The
interaction of this field with the crystal in the dipole
approximation is described by $H_{int}=-(e/mc){\bf p\cdot A}(t)$,
where $e (m)$ is the charge(mass) of an electron, $c$ is the speed
of light and ${\bf A}(t)$ is the vector potential associated with
the field. In the ground state $ |0\rangle $, electrons occupy
states up to the Fermi level. The wavefunction of the perturbed
system at time $t$ is then written as $|\Psi (t)\rangle
=c_{0}(t)|0\rangle + \mathop{\displaystyle \sum } _{{\bf
k}}c_{{\bf k}}(t{\bf )}|eh{\bf k}\rangle $, where $|eh{\bf
k}\rangle $, denotes the state where an electron ($e$) and hole
($h$) are generated at point ${\bf k}$ (see Fig. 1). Time
dependent perturbation theory is used to evaluate $c_{{\bf
k}}(t{\bf )}$ and hence to describe the rate of change in the
expectation value of a single particle operator $\theta$ in the
independent particle approximation,
\begin{equation}
\frac{\partial \langle \theta\rangle }{\partial t}=
\mathop{\displaystyle \sum } _{{\bf k}}\langle eh{\bf k}|{\theta
}|eh{\bf k}\rangle \frac{
\partial \left( c_{{\bf k}}(t{\bf )}c_{{\bf k}}^{*}(t{\bf )}\right) }{
\partial t}. \label{general}
\end{equation}
This leads to the evaluation of injection rates of interest,
including that for the density of flipped spins and for the spin
current density \cite{Review}. The former can be written as
\begin{equation}
\frac{\partial n}{\partial t}=\xi ^{abcd}(\omega
_{1},\Omega)E^{a}{\bf ( }\omega _{1}{\bf )}E^{b*}{\bf (}\omega
_{2}{\bf )}E^{c*}{\bf (}\omega _{1} {\bf )}E^{d}{\bf (}\omega
_{2}{\bf )}, \label{ndot}
\end{equation}
\newline
where $\xi ^{abcd}(\omega_1, \Omega)=V^{-1} \mathop{\displaystyle
\sum } _{{\bf k}}\Gamma ^{abcd}({\bf k})\delta (\Omega-\omega
_{eh}({\bf k}))$ and
\begin{eqnarray}
\Gamma ^{abcd}({\bf k})&=&\left( \frac{\sqrt{2\pi }e^{2}}{\hbar
^{2}\omega _{2}\omega _{1}}\right) ^{2}\left\{ \left[
\mathop{\displaystyle \sum } \limits_{n}\frac{v_{nh}^{a}({\bf
k})v_{en}^{b}({\bf k})}{\omega _{1}+\omega _{hn}({\bf k})} \right.\right.
\nonumber \\
&&\hspace{-0.7cm}\left.\left. -\mathop{\displaystyle \sum }
\limits_{n}\frac{v_{en}^{a}({\bf k})v_{nh}^{b}({\bf k})}{\omega
_{1}-\omega _{en}({\bf k})}\right] \times (a\rightarrow
c,b\rightarrow d)^{*}\right\}. \label{gamma}
\end{eqnarray}
Here $\omega _{nm}=\omega _{n}-\omega _{m}$, $V$ is the
normalization volume, {\it a,b,c} and {\it d} are the Cartesian
indices and ${\bf v} _{nm}({\bf k})$ is the velocity matrix
element between single particle states $n$ and $m$ at ${\bf k}$.
Repeated indices are summed over. To quantify the spin current
density ${J}_{ab}$, we use the conventional
\cite{Rashba,Erlingsson} single particle spin current operator
$j_{ab}=(1/2)\left(v^{a}S^{b}+S^{b}v^{a}\right) $, where
${\bf{S}}=\hbar {\bm \sigma }/2$ and ${\bm \sigma }=\left( \sigma
^{x},\sigma ^{y},\sigma ^{z}\right) $ are the Pauli matrices. The
spin current density injection rate is
\begin{equation}
\frac{\partial J_{ab}}{\partial t}=\mu
^{abcdfg}(\omega_1,\Omega)E^{c}{\bf (}\omega _{1}{\bf )
}E^{d*}{\bf (}\omega _{2}{\bf )}E^{f*}{\bf (}\omega _{1}{\bf
)}E^{g}{\bf (} \omega _{2}{\bf )}, \label{Jdot}
\end{equation}
where $\mu ^{abcdfg}(\omega_1,\Omega)=V^{-1} \mathop{\displaystyle
\sum } _{eh{\bf k}}\delta j_{ab}\Gamma^{cdfg}({\bf k})\delta
(\Omega-\omega _{eh}({\bf k}))$ and $\delta j_{ab}=\langle e{\bf
k}| {j}_{ab}|e{\bf k}\rangle -\langle h{\bf k}|{j}_{ab}|h{\bf k}
\rangle $ is the spin current contribution per spin flip
\cite{footnote1}. Both the fourth rank tensor $\xi ^{abcd}=\xi
^{cdab}$ and sixth rank pseudotensor $\mu ^{abcdfg}=\mu ^{abfgcd}$
can be shown to be real by using time reversal symmetry; the
$T_{d}$ symmetry group of the crystal can be used to deduce the
non-zero and independent tensor components. To calculate the
tensors, we describe the spin splitting of the conduction states
by the Dresselhaus Hamiltonian in the $\Gamma $ point basis
$\left\{ |1/2,1/2\rangle ,|1/2,-1/2\rangle \right\}$
\begin{equation}
H=\frac{\hbar ^{2}k^{2}}{2m_{c}}+\alpha _{D}{\bm \kappa }\cdot
{\bm \sigma ,} \label{H}
\end{equation}
where $\kappa_{x}=k_{x}(k_{y}^{2}-k_{z}^{2})$ and $\kappa _{y,z}$
are cyclic permutation of $\kappa _{x}$, $\alpha _{D}=27.6$ $\rm
eV\AA^3$ is the Dresselhaus parameter and $m_{c}=0.067m$ is the
electron effective mass \cite{Dresselhaus,Winkler}. The spin
splitting at wavevector $ {\bf k}$ is $2\alpha _{D}\kappa $. The
valence states of the semiconductor serve as the intermediate
virtual states $n$ in Eq. (\ref{gamma}). The heavy and light
valence states are modelled using the isotropic $4\times 4$
Luttinger-Kohn Hamiltonian, with the Luttinger parameters $\gamma
_{1}=6.85$ and $\gamma =\left( \gamma _{2}+\gamma _{3}\right)
/2=2.5$. The split-off band is described by a diagonal $ 2\times
2$ Hamiltonian, with the effective mass taken to be $0.154m$
\cite{Loehr}. A band gap energy of $E_{g}=1.51$ {\rm eV} and a
split-off energy of $0.34$ {\rm eV} are used. The Fermi energy of
the electrons in the conduction band is given by
$E_{F}=E_{g}+\hbar ^{2}k_{F}^{2}/2m_{c}$, where $k_{F}=(3\pi
^{2}N)^{1/3}$ for the carrier concentration $N$; we use a
concentration of $N=10^{18}$ {\rm cm}$^{-3}$ in the calculations.
The speed of the injected carriers is given by $v_{F}=\hbar
k_{F}/m_{c}\approx 500$ {\rm km/s}. The intraband velocity matrix
elements are evaluated using $ {{\bf v}}=\hbar ^{-1}\partial
{H}({\bf k})/\partial {\bf k}$ as the velocity operator. We use a
Kane parameter $P=(\hbar /m)\langle S|{p}_{x}|X\rangle =10.49$
{\rm eV\AA\ }\cite{Loehr} and numerically evaluate the tensor
components.

Spin orbit coupling splits the Fermi-surface into two surfaces.
The fraction of the electrons that are contained in the region
between the two surfaces is given by $\approx 2m_{c}\alpha
k_{F}/\hbar ^{2}$ and is about $1\%$ of the free carriers
available in the system at $N=10^{18}$ {\rm cm}$^{-3}$. Only this
fraction of the free carriers in the system is subject to Raman
scattering. This sets a limit of $\approx 10^{-2}N$ on the
concentration of carriers that can contribute to the injected
current.

Spin flip Raman scattering has typically been studied using
orthogonally polarized fields ${\bf E}(\omega _{1,2})$
\cite{LightScattering}. Hereafter we use the following definition
of the coordinates, $ \hat{x}\equiv [001],$ $\hat{y}\equiv
[1\overline{1}0]$ and $\hat{z}\equiv [110]$ and consider more
general fields normally incident on a crystal with a $[110]$
growth axis
\begin{equation}
{\bf E}(\omega _{1,2})=E_{1,2}\beta _{1,2}\left( \cos (\rho
_{1,2}^{x}) \hat{x}+i\cos (\rho _{1,2}^{y}){\hat{y}}\right),
\label{fields}
\end{equation}
for phases $\left( \rho _{1}^{x},\rho _{1}^{y},\rho _{2}^{x},\rho
_{2}^{y}\right) $, and normalization factors $\beta _{1,2}$. This
allows the most general analysis of the fields polarizations. We
find the spin-flip and spin current density injection rates
\begin{eqnarray}
\frac{\partial n}{\partial t} &=&g\cdot A_{\varepsilon _{1}}(\Omega
)I_{1}I_{2},  \nonumber \\
\frac{\partial J_{yz}}{\partial t} &=&g\cdot \frac{\hbar }{2}\left[
B_{\varepsilon _{1}}(\Omega )+C_{\varepsilon _{1}}(\Omega )\right]
I_{1}I_{2},  \label{results} \\
\frac{\partial J_{zy}}{\partial t} &=&g\cdot \frac{\hbar }{2}\left[
B_{\varepsilon _{1}}(\Omega )-C_{\varepsilon _{1}}(\Omega )\right]
I_{1}I_{2},  \nonumber
\end{eqnarray}
where $g=\left( \beta _{1}\beta _{2}\right) ^{2}\left( \cos (\rho
_{1}^{y})\cos (\rho _{2}^{x})+\cos (\rho _{1}^{x})\cos (\rho
_{2}^{y})\right) ^{2}$, and $I_{1,2}$ are the incident intensity
of the $\varepsilon _{1,2}=\hbar \omega _{1,2}$ components of the
fields. Here $A_{\varepsilon _{1}}$, $B_{\varepsilon _{1}}$and $
C_{\varepsilon _{1}}$ are linear combination of the tensor
components $\xi ^{abcd}$ and $\mu ^{abcdfg}$. The mismatch in the
dielectric function $\epsilon$ across the interface leads to a
ratio $ 2/(\sqrt{\epsilon}+1)$ of the electric fields in the
crystal and air. Two pure spin currents are injected in the
crystal, one injected along $y=[1 \overline{1}0]$ with spins
parallel to $z=[110]$ and another injected along $z=[110]$ with
spins pointing parallel to $y=[1 \overline{1}0]$ (Fig. 1). Clearly
for either right (or left) circularly (e.g. $\beta _{1}=\beta
_{2}=1/\sqrt{2}$, $\rho _{1}^{x}=\rho _{2}^{x}=0$, $ \rho
_{1}^{y}=\rho _{2}^{y}=0 (\rm{or}$ $\pi )$) and cross-linearly
(e.g. $\beta _{1}=\beta _{2}=1$, $\rho _{1}^{x}=\rho _{2}^{y}=0$,
$\rho _{1}^{y}=\rho _{2}^{x}=\pi /2$) polarized fields, the same
spin currents are injected. Moreover, for co-linearly polarized
fields, where $g=0$, no spin-flips and hence no pure spin current
can be induced.

First we consider the case where the photon energy
$E_{\rm{I}}=\hbar \omega _{1}=0.754$ {\rm eV }is close $E_g/2$.
All two photon absorption from the valence to the conduction bands
is avoided in this scenario. In Fig. 2, we present $A_{\rm{I}}$,
$B_{\rm{I}}$ and $C_{\rm{I}}$. The role of the resonance
enhancement can be investigated by using optical fields of energy
$E_{\rm{II}}=\hbar \omega _{1}=1.49$ {\rm eV} close to the band
gap. Transitions from the partially filled conduction band to the
higher bands are avoided at this energy, but two photon
excitations from the valence states to conduction states are
allowed. Nonetheless these transitions are away from the $\Gamma $
point and should not contribute to the injected pure spin current
from the spin flip Raman processes we discuss in this letter. The
two photon absorption is estimated to be an order of magnitude
smaller than the spin-flip injection rate discussed here
\cite{Atanasov}, and excitons are not generated. The carrier and
spin current density injection rates characterized by
$A_{\rm{II}}$, $B_{\rm{II}}$ and $C_{\rm{II}}$ are presented in
Fig. 2. The sharp features in the results, shown in Fig. 2,
correspond to the van Hove singularities in the joint density of
states for the spin-flip transitions. The van Hove singularity is
due to the vanishing derivative in the spin splitting, $2\alpha
_{D}\nabla _{{\bf k}}\kappa $ along high symmetry directions, and
an analysis of the Dresselhaus Hamiltonian shows that the
resulting density of states increases as $\ln(1/\left| \Omega
-\Omega _{vH }\right| ),$ where $\Omega _{vH }$ is the position of
the singularity.
\begin{figure}
\includegraphics[width=6.0cm]{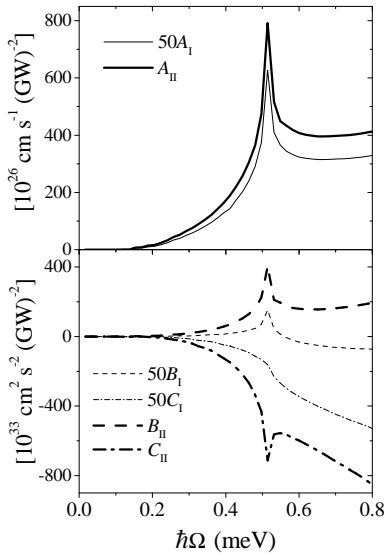}
\caption{The thin lines show the coefficients, $A_{\rm{I}}$,
$B_{\rm{I}}$ and $C_{\rm{I}}$ Eq. (\ref{results}) for
$\Omega=\omega_{1}-\omega_{2}$ and $\hbar \omega_1 = 0.754$ eV,
about half the band gap. The thick lines show the coefficients,
$A_{\rm{II}}$, $B_{\rm{II}}$ and $C_{\rm{II}}$ Eq. (\ref{results})
for $\hbar \omega_1 = 1.49$ eV, about $20$ {\rm meV} below the
band gap.}
\end{figure}
To this point we have discussed the injection rate of pure spin
currents by stimulated Raman scattering using two nominally
continuous wave (CW) beams at distinct frequencies $\omega_2$ and
$\omega_1 > \omega_2$. We now turn to the ultrafast pulse regime,
and stress that a pure spin current can be injected through the
Raman effect by a single pulse, with the scattering now arising
from the spectral width of the pulse. For a pulse with a centre
frequency $\omega_{0}$ and width $\Delta \omega$, if we have
$\omega_{0} \gg \Delta \omega$ and $\omega_{0} \gg
\Omega_{\rm{max}}$, where $\hbar \Omega_{\rm{max}}$ is the maximum
spin splitting, the straightforward generalization of the
calculation in Eqs. (\ref{ndot}) and (\ref{Jdot}) leads to the
injection of densities of spin flips and spin current given by
\begin{equation}
\Delta n =\int\limits_{0}^{\Omega _{\max }}\frac{d\Omega
}{2\pi}\xi ^{abcd}(\omega _{0},\Omega )G^{abcd}(\Omega ),
\label{pulse_result1}
\end{equation}
\begin{equation}
\Delta J_{ab} =\int\limits_{0}^{\Omega _{\max }}\frac{d\Omega
}{2\pi }\mu ^{abcdfg}(\omega _{0},\Omega )G^{cdfg}(\Omega ),
\label{pulse_result2}
\end{equation}
where the field correlation function for ${\bf
E}(t)=(2\pi)^{-1}\int_{-\infty}^{\infty}d\omega{\bf E}(\omega
)e^{-i\omega t}$ is
\begin{eqnarray}
G^{abcd}(\Omega ) &=& \\
&&\hspace{-2cm}\int\limits_{0}^{\infty }\frac{d\omega _{1}d\omega
_{1}^{\prime }}{\left( 2\pi \right) ^{2}}E^{a}(\omega
_{1})E^{b*}(\omega _{1}-\Omega )E^{c*}(\omega _{1}^{\prime
})E^{d}(\omega _{1}^{\prime }-\Omega ). \nonumber \label{G}
\end{eqnarray}
We consider a (right or left) circularly polarized short pulse of
the form $E(t) = 2 E_{0} \exp (-t^{2}/\tau ^{2})\cos(\omega
_{0}t)$, with fluence (pulse energy per unit area) $F$. A short
pulse \cite{Shank} ($\tau \Omega _{\rm{\max} } \ll 1 $ and $\tau
\ll \mu m_c/|e|$, where $\mu$ is the mobility) essentially leads
to the generation of the spin current in the ballistic limit. For
$\hbar \omega _{0}=0.754$ {\rm eV} (corresponding to $A_{\rm{I}}$,
$B_{\rm{I}}$ and $C_{\rm{I}}$ in Fig. 2), the density of spin
flips per pulse is $\Delta n_{\rm{I}}\approx (1.8\times 10^{20}
{\rm{ cm}/\rm \it{J}^{\rm{2}}})F^{2}$, which gives $\Delta
n_{\rm{I}}\approx 10^{15} \rm{cm^{-3}}$ for an experimentally
accessible fluence of $F=3\times 10^{-3}$ {\rm
\it{J}/\rm{cm}}$^{2}$. In a scenario where $\hbar \omega
_{0}=1.49$ {\rm eV} (corresponding to $A_{\rm{II}}$, $B_{\rm{II}}$
and $C_{\rm{II}}$ in Fig. 2) the spin flips density due to Raman
scattering is $\Delta n_{\rm{II}}\approx (1.\,2\times 10^{22}
{\rm{cm}/\rm \it{J}^{\rm{2}}})F^{2}$; a spin flip density of
$\Delta n_{\rm{II}}\approx 10^{15} \rm{cm^{-3}}$ can be achieved
for a fluence of $F=3\times 10^{-4}$ {\rm \it{J}/\rm{cm}}$^{2}$,
which reflects a resonance enhancement of the process by two
orders of magnitude.

Note that the injection in this process is proportional to
$F^{2}$, whereas from Eq. (\ref{results}) for nominal CW
irradiation one might expect a scaling with $I^2_{\rm{max}}\tau
\propto F^2/\tau$ where $I_{\rm{max}}$ is the maximum intensity in
the pulse. The reason is that, for $\Omega_{\rm{max}}\tau \ll 1$,
a factor $\Omega_{\rm{max}}\tau \approx \Omega_{\rm{max}}/\Delta
\omega$ appears in the final result following from Eqs.
(\ref{pulse_result1}) and (\ref{pulse_result2}), reflecting the
fact that only a fraction $\Omega_{\rm{max}}/\Delta \omega$ of the
pulse bandwidth can contribute to the Raman scattering.

With each pulse a number of carriers $\approx \Delta n \times S
\eta$ are excited, where $S$ is the laser spot area and $\eta$ is
the penetration depth. In earlier experiments
\cite{Hubneretal,Stevensetal} involving interband transitions, a
large number of carriers are available in the valence states, and
typically $\approx (10^{18}{\rm cm^{-3}}) S \eta$, where $\eta
\approx 1$ $\mu m$, carriers can contribute to the spin currents.
On the other hand, in Raman injected spin currents the penetration
depth is much larger than that for interband transitions, with a
limit $\eta^{\prime} \approx 10^3 \mu$m set by the Drude
absorption due to free carriers. This allows a large $\approx
(10^{15} {\rm cm^{-3}}) S \eta^{\prime}$ number of spin flips in
the sample, which will then contribute to the injected pure spin
current.

Pump and probe schemes have been used to experimentally observe
injected pure spin currents \cite{Stevensetal}. An experimentally
relevant parameter in these schemes is the separation between the
opposite spins after momentum relaxation. This distance can be
estimated by $d^{a}=\left( 2\mu m_{c}/|e|\hbar \right)
\sum_{b}\left| \stackrel{\cdot }{J}_{ab}\right| /\stackrel{\cdot
}{n} $ \cite{footnote2,mobilityRef}. For circularly polarized as
well as cross linearly polarized fields, a spin separation of
$d^{y}\approx 18$ nm for $\mu \approx 5\times 10^{3}$
cm$^{2}$/(Vs) is predicted, a distance which is within
experimental capabilities \cite{Stevensetal}. Measuring a pure
spin current by using the anomalous spin Hall effect arising due
to skew scattering by impurities \cite{Abakumov72,Nozieres} has
also been proposed \cite{Shermanetal}. The spin current $J_{ab}$
causes a spin-Hall bias $ V_{sH} $ along the direction
perpendicular to the electron propagation. This bias is estimated
as $V_{sH}\approx \tan (\theta _{sH})V_{{\rm eff}}$, where $\theta
_{sH}$ is the spin-Hall angle and $V_{{\rm eff}}$ is the effective
bias that would cause a current density $eJ_{ab}/\hbar$; it is of
the order of $e\Delta nv_{F}$, where $\Delta n$ is the number of
carriers that form the pure spin current. Therefore $V_{{\rm
eff}}\approx L\left( \Delta n/N\right) v_{F}/\mu $, where
$L\approx 10$ $\mu$m is the lateral size of the system, which is
assumed to be approximately equal to the spot size. The ratio
$\Delta n/N$ is estimated as $\zeta \alpha _{D}k_{F}^{3}/E_{F}$,
where $\zeta $ is the fraction of Raman accessible electrons that
undergo a spin flip. With the above mobility we obtain: $V_{{\rm
eff} }/L\approx 10\zeta $ ${\rm V/cm}$. The above analysis shows $
\zeta$ to be about 0.1. With the angle $\theta _{sH}$ estimated by
Abakumov and Yassievich \cite{Abakumov72} for GaAs at low
temperatures as $ \theta _{sH}\approx 10^{-3}$, we obtain an upper
estimate of $V_{sH}\approx 10^{-6}$ ${\rm V}$.

In conclusion, stimulated spin-flip Raman scattering can be used
to inject spin currents in doped semiconductors with spin
splitting. We have considered the case of an {\it n}-doped GaAs
bulk crystal, and showed that the injected spin current is a pure
spin current comprised of spins with opposite orientations
travelling in opposite directions. This pure spin current should
be experimentally accessible, and can be achieved either with two
continuous waves contributing to the Raman scattering, or by using
a single pulse. The scheme is all-optical but, unlike other
optical schemes involving the excitation of electron-hole pairs,
here there is minimal energy deposited in the crystal. The system
is thus only slightly perturbed from equilibrium, and the
description of transport and the design of device structures
should be easier than in scenarios employing other optical
schemes. High mobility quantum wells should provide ideal testing
grounds for the general proposal discussed in this letter.

A.N. acknowledges support from an OGS. This work was supported in
part by the NSERC (Canada) and DARPA. We thank D. Luxat, P.A.
Marsden, J. H\"{u}bner, H.M. van Driel, and A.L. Smirl for useful
discussions.


\begin{references}
\bibitem{footnote} While we only consider the Stokes Raman process,
the thermal spreading of the Fermi distribution in both spin-split
bands will allow for anti-Stokes processes. However, the
combination of the Stokes and anti-Stokes contributions can be
shown to be weakly temperature dependent.
\bibitem{footnote1} In this letter we neglect the uncertainty in the
    momentum dependent spin-splitting; which is allowed since the momentum
    is a well-defined quantum number: the product
    $k_F \l \approx 10$, where $k_F$ is the Fermi
    wavevector and $\l$ is the mean free path. We also assume that
    the inelastic broadening of the states arising due to inelastic scatterings
    is small compared to $\alpha_{D}\kappa$
    due to low $T$ even at $k_{B}T>\alpha_D\kappa$ \cite{Jusserandetal}. The
    role of relaxation processes in the Raman injection of spin
    current is, however, an interesting problem and requires a separate investigation.
\bibitem{footnote2} The experimentally observed mobility at the
assumed density and at 300 K is $\approx$ 3000 $\rm{cm}^2$/(Vs).
We take the mobility 5000 $\rm{cm}^2$/(Vs) for the low
temperatures considered here.
\end{references}
\end{document}